**Acoustic toroidal vortices with programmable links and knots**

Shuai Liu[1,2], Xiang-Yuan Xu[2], Hao Ge[1,*], Yijie Shen[3,4,†], Yan-Feng Chen[1,2], and Ming-Hui Lu[1,2,5 ‡]

*[1]National Laboratory of Solid State Microstructures & School of Advanced Manufacturing Engineering, Nanjing University, Suzhou, Jiangsu 215163, China*

*[2] College of Engineering and Applied Sciences & Collaborative Innovation Center of Advanced Microstructures, Nanjing University, Nanjing, Jiangsu 210093, China*

*[3] Centre for Disruptive Photonic Technologies, School of Physical and Mathematical Sciences, Nanyang Technological University, Singapore 637371, Singapore*

*[4] School of Electrical and Electronic Engineering, Nanyang Technological University, Singapore 639798, Singapore*

*[5] Jiangsu Key Laboratory of Artificial Functional Materials & Jiangsu Physical Science Research Center, Nanjing University, Nanjing, Jiangsu 210093, China*

*Correspondence should be addressed to: haoge@nju.edu.cn, yijie.shen@ntu.edu.sg or luminghui@nju.edu.cn*

## Abstract

Toroidal vortices are three-dimensional torus-shaped wave structures characterized by phase circulation around a closed vortex line. Their toroidal geometry provides a natural foundation for constructing linked and knotted wave structures. Here we experimentally synthesize scalar acoustic toroidal vortices using a programmable circular phased array. Full spatiotemporal measurements directly resolve the toroidal envelope, the closed phase-singularity ring, the associated poloidal phase winding, and the free-space evolution of the wave packet. By introducing an independently controlled phase winding along the toroidal cycle, we realize scalar acoustic hopfions and directly reconstruct their three-dimensional equiphase fibers from the measured complex pressure field. Varying the poloidal and toroidal winding numbers controls the phase-fiber geometry, linking, and connectivity, yielding a Hopf link, a multicomponent torus link, and a trefoil knot. These results provide direct experimental access to the geometry, propagation dynamics, and phase-fiber topology of scalar toroidal wave fields, establishing a reconfigurable acoustic platform for linked and knotted wave structures.

## Introduction

Toroidal vortices, commonly known as vortex rings, are three-dimensional ring-shaped structures in which vortical motion is organized around a closed vortex line. They are widely encountered in fluid dynamics [1,2], appearing in phenomena ranging from smoke and bubble rings to cardiovascular flow [3], cloud dynamics, and biological transport [4]. Their closed-loop geometry gives rise to characteristic self-induced motion and supports a wide range of dynamical behaviors, including shape deformation, instability, interaction, and reconnection [5–8]. Related toroidal structures have also been explored in magnetic media [9,10], Bose–Einstein condensates [11], and plasmonic systems [12].

In wave physics, toroidal wave structures have been extensively studied in electromagnetism and optics [13], where both vector and scalar realizations have emerged. Vector toroidal pulses are characterized by coupled electric and magnetic fields arranged in a toroidal geometry [14–19], whereas scalar toroidal vortices are phase-vortex wave packets whose spatiotemporal vortex line forms a closed ring [20–27]. In each local radial–temporal cross section normal to the ring, the phase circulates around the singularity, giving rise to a transverse orbital angular momentum density tangent to the ring [28–31]. Hybrid electromagnetic toroidal vortices combining scalar phase singularities with vector toroidal field textures have also been demonstrated [32]. Beyond the elementary scalar toroidal vortex, a toroidal surface possesses two independent noncontractible cycles: the poloidal cycle around the minor circumference and the toroidal cycle around the major circumference [33–35]. Introducing an additional phase winding along the toroidal cycle controls how equal-phase points in successive radial–temporal sections connect around the closed ring. When both winding numbers are nonzero, the constant-phase curves wind around the two cycles of the torus, forming closed phase fibers on toroidal iso-amplitude surfaces. This construction underlies scalar optical hopfions with Hopf-linked and torus-knot phase fibers [33,34,36,37].

In parallel, acoustic vortex physics has expanded from conventional spatial vortices [38–41] to spatiotemporal vortices supporting transverse orbital angular momentum [42–45]. More recently, spatiotemporal acoustic vortex rings have been generated through scattering-induced spatiotemporal coupling [46]. Separately, spatial acoustic vortex lines have been shaped into knots and links in three-dimensional real space [47]. These are purely spatial singularity structures, distinct from the toroidal wave packets considered here, which are defined over two transverse spatial coordinates and time. Toroidal acoustic states carrying two winding indices have also been realized through highly degenerate flatband modes in three-dimensional acoustic crystals [48], but are confined to the structured lattice rather than propagating freely. These developments establish a broad foundation for toroidal, linked, and knotted acoustic fields and motivate their investigation in a freely propagating setting with full complex-field reconstruction and programmable control of topology.

Here, we synthesize freely propagating scalar acoustic toroidal vortices with a programmable circular phased array. Spatiotemporal sampling reconstructs the complex pressure field over two transverse spatial coordinates and time, directly revealing the toroidal envelope, the closed phase-singularity ring, the local poloidal phase winding, and the evolution of these features during free propagation. We then realize scalar acoustic hopfions by introducing an independently controlled phase winding along the toroidal cycle and reconstruct the resulting phase fibers on iso-amplitude tori from the measured fields. Varying the poloidal and toroidal winding numbers produces fiber configurations with distinct winding geometries, linking relations, and numbers of connected components, including a Hopf link, a multicomponent torus link, and a trefoil knot. These results show that freely propagating acoustic toroidal vortices can support programmable phase-fiber topology encompassing linking, connectivity, and knotting.

## Theoretical analysis and experimental setup

We begin from the scalar acoustic wave equation

$$\left(\nabla^2 - \frac{1}{c^2}\frac{\partial^2}{\partial t^2}\right)p(\boldsymbol{r},t) = 0, \tag{1}$$

where $p(\boldsymbol{r},t)$ is the acoustic pressure and $c$ is the speed of sound. The target field is specified on a reference plane $z = 0$ in cylindrical coordinates $r = \sqrt{x^2 + y^2}$ and $\varphi = \arctan\left(\frac{y}{x}\right)$. To place the temporal coordinate on the same dimensional footing as the transverse coordinates, we introduce $\xi = t/s_t$, where $s_t$ is a tunable space-time scaling parameter. The local polar coordinates in a radial-time, or poloidal, section are $\rho = \sqrt{(r - r_0)^2 + \xi^2}$ and $\theta = \arctan\left(\frac{\xi}{r - r_0}\right)$, with $r_0$ the major radius of the torus. A family of scalar toroidal wave packets can then be written as

$$p_{env}(x,y,\xi) \propto \left(\frac{\rho}{w}\right)^{|\ell_p|} \exp\left(-\frac{\rho^2}{2w^2}\right) \exp\left(il_{\mathrm{tor}}\varphi - il_{\mathrm{pol}}\theta\right), \tag{2}$$

where $w$ controls the minor radius of the intensity torus. The integers $l_{\mathrm{pol}}$ and $l_{\mathrm{tor}}$ denote the poloidal and toroidal phase winding numbers, respectively, and are defined by

$$l_{\mathrm{pol}} = \frac{1}{2\pi}\oint_{C_{\mathrm{pol}}} \nabla\Phi \cdot d\boldsymbol{r}, \tag{3}$$

$$l_{\mathrm{tor}} = \frac{1}{2\pi}\oint_{C_{\mathrm{tor}}} \nabla\Phi \cdot d\boldsymbol{r}. \tag{4}$$

Here, $C_{\mathrm{pol}}$ and $C_{\mathrm{tor}}$ are closed contours taken along the poloidal and toroidal directions, as illustrated in Figs. 1(a) and 1(b). $C_{\mathrm{pol}}$ encircles the minor cross section in the direction of decreasing $\theta$, whereas $C_{\mathrm{tor}}$ follows the major circumference in the direction of increasing $\varphi$. As shown in Fig. 1(a), for $l_{\mathrm{tor}} = 0$ and $l_{\mathrm{pol}} \neq 0$, the phase winds only along the poloidal cycle, producing a scalar acoustic toroidal vortex with a closed phase-singularity ring. Figure 1(b) shows the case with an additional nonzero $l_{\mathrm{tor}}$, which introduces a phase winding along the toroidal cycle and yields a scalar acoustic hopfion with coupled poloidal and toroidal phase windings.

To synthesize the target wave packet described by Eq. (2), we first modulate its complex envelope with a carrier at angular frequency $\omega_0$, such that $p(x, y, 0, t) = p_{env}(x, y, \xi)e^{-i\omega_0 t}$. The corresponding spatiotemporal spectrum is then obtained by the three-dimensional Fourier transform

$$P\left(k_x, k_y, \omega\right) = \iiint p(x, y, 0, t)e^{-i(k_x x + k_y y) + i\omega t}\, dx\, dy\, dt\,. \tag{5}$$

We sample the spectrum within $k_x, k_y \in [-0.6k_0, 0.6k_0]$ and $\omega \in [0.4\omega_0, 1.6\omega_0]$, and retain 2045 propagating modes satisfying $k_x^2 + k_y^2 \leq (\omega/c)^2$. The phased array simultaneously generates these modes with the complex weights prescribed by the sampled spectrum. Their superposition reconstructs the target spatiotemporal wave packet.

The experimental scheme for synthesizing these spatiotemporal toroidal acoustic fields is illustrated in Fig. 1(c), where a circular acoustic phased array generates a toroidal wave packet with a closed ring-shaped phase singularity. The multichannel driving signals, constructed from the superposition of the selected plane-wave components, are calculated on a computer and delivered through a synchronized sound card to the circular acoustic phased array, which radiates the programmed broadband acoustic field. A microphone scans the $x - y$ measurement plane over a $72 \times 72$ grid with a spatial sampling interval of $1\,\mathrm{cm}$. At each position, the time-domain pressure waveform is recorded at a sampling rate of $51.2\,\mathrm{kHz}$ and synchronously collected by the data-acquisition module for reconstruction of the complex spatiotemporal acoustic field [see Supplemental Material [49], Sec. III ].

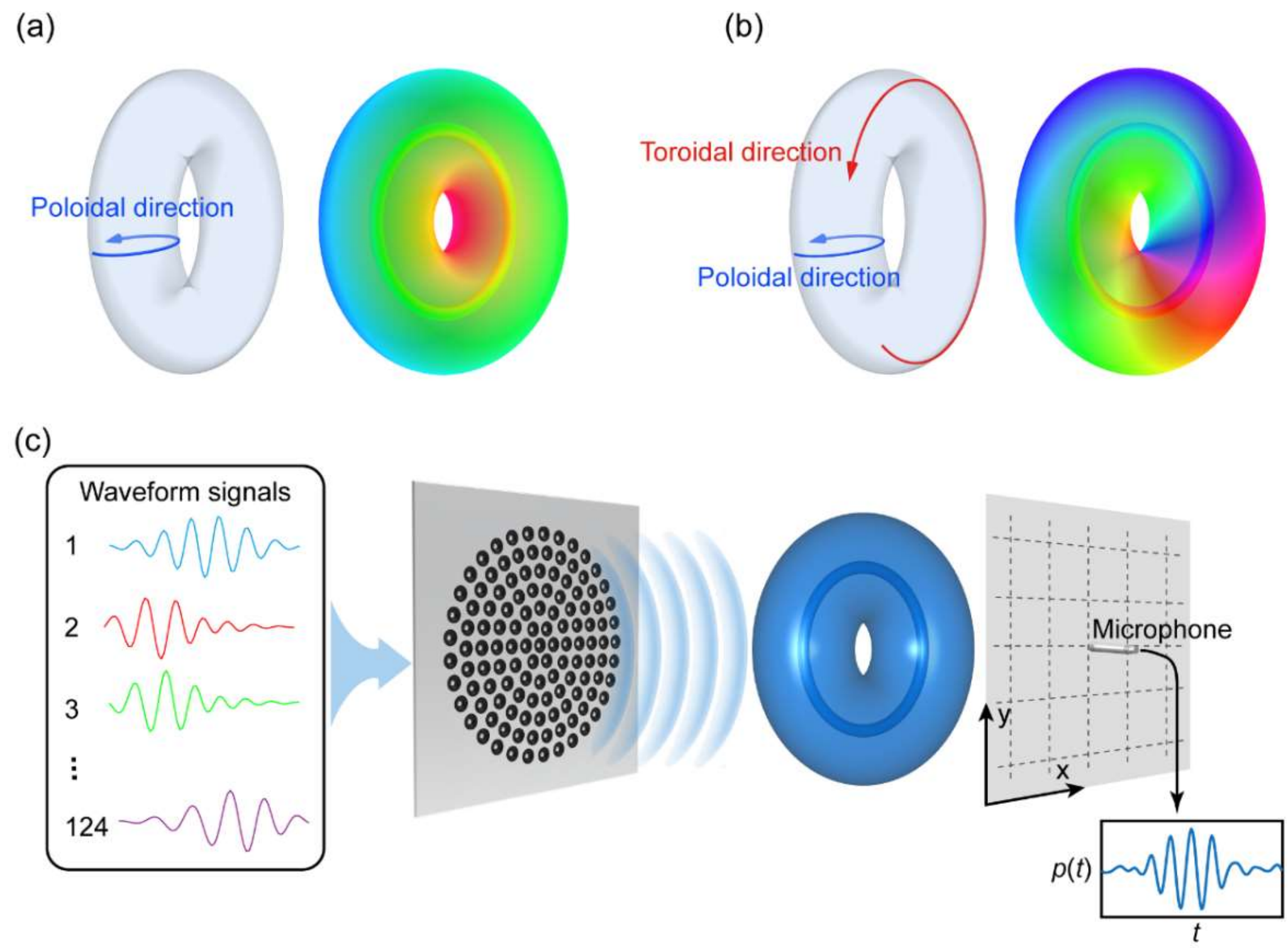

**FIG. 1. Concept and generation of scalar acoustic toroidal vortices and hopfions.** (a) Schematic of a scalar acoustic toroidal vortex with phase winding along the poloidal direction. The color on the toroidal iso-amplitude surface represents the phase. (b) Scalar acoustic hopfion obtained by introducing an additional phase winding along the toroidal direction. The blue and red arrows indicate the poloidal and toroidal winding directions, respectively. (c) Experimental setup. A 124-channel circular phased array is driven by independently programmed waveform signals to synthesize the desired toroidal wave packet. A microphone scans the transverse $x-y$ plane and records the time-domain acoustic pressure signal at each position.

**Observation and free-space evolution of scalar acoustic toroidal vortices**

We first consider the fundamental scalar acoustic toroidal vortex with $l_{\mathrm{tor}} = 0$ and $l_{\mathrm{pol}} = 1$, for which the phase winds once around the poloidal direction while remaining uniform along the toroidal direction. Figure 2 compares the calculated and experimentally reconstructed fields at the reference plane $z = 0$. The theoretical iso-amplitude surface in Fig. 2(a) forms a continuous toroidal shell surrounding a closed ring-shaped phase singularity. The perspective and front views reveal the annular geometry in the transverse plane, whereas the side and top views show that the low-pressure core extends as a closed circular line in the three-dimensional $(x, y, t)$ domain. Experimentally, the reconstructed field in Fig. 2(b) forms a continuous toroidal envelope surrounding a closed ring-shaped phase singularity, directly revealing the three-dimensional closed-vortex geometry.

The phase structure is further resolved through two orthogonal radial-time sections. Figures 2(c) and 2(d) show the theoretical and measured complex pressure distributions in the $x - t$ plane at $y = 0$ and in the $y - t$ plane at $x = 0$, respectively, where brightness represents the normalized pressure amplitude and hue encodes the phase. In each section, the toroidal singular ring is intersected at two points, appearing as two isolated pressure zeros. Around each zero, the phase undergoes a full $2\pi$ circulation, confirming a first-order spatiotemporal vortex. The two singularities in each diametric section are therefore not independent defects, but intersections with the same closed vortex line. The measured phase circulation around the two intersections directly identifies the local poloidal winding associated with the same closed vortex line.

We next investigate the free-space evolution of the toroidal vortex. The theoretical propagation is calculated using a spatiotemporal angular-spectrum method, in which the field at the reference plane is decomposed into plane-wave components $P(k_x, k_y, \omega)$. Each component acquires a propagation phase determined by the longitudinal wave-vector component

$$k_z = \sqrt{\left(\frac{\omega}{c}\right)^2 - k_x^2 - k_y^2}, \tag{6}$$

which follows from the acoustic dispersion relation in air, where $c$ denotes the speed of sound. After each component accumulates the propagation phase $\exp(ik_z z)$, the field at each propagation distance is obtained by summing all plane-wave components. Figure 3 presents the calculated and measured fields at propagation distances $z = 0, 0.2,$ and $0.4$ m. At the reference plane, the field exhibits a compact toroidal envelope with an approximately circular singular ring. During propagation, the phase accumulated by each plane-wave component depends on both its frequency and transverse wave vector, progressively changing the relative phases imposed at the reference plane. Over the present bandwidth, air provides negligible temporal material dispersion, so the radial and temporal profiles no longer evolve in the same manner. Each local poloidal section consequently broadens and distorts, and the collective deformation of these sections around the closed ring reshapes the entire toroidal wave packet. At $z = 0.2$ m, both the closed singular ring and the surrounding toroidal iso-amplitude surface undergo noticeable deformation, indicating an overall distortion of the spatiotemporal wave packet. At $z = 0.4$ m, the deformation becomes substantially stronger: the original toroidal envelope is stretched and partially fragmented, and the radial-time intensity pattern evolves into a broadened central structure accompanied by weaker side lobes.

The evolution is more clearly seen in the radial-time sections shown in Fig. 3(c). At $z = 0$, both theory and experiment exhibit two well-separated high-amplitude regions corresponding to the two intersections of the toroidal wave packet with the selected radial-time plane. At $z = 0.2$ m, these sectional profiles broaden and become noticeably distorted as the singular ring and the surrounding wave-packet envelope deform during propagation. At $z = 0.4$ m, the deformation becomes more pronounced, and the two initially separated sections evolve into a broadened and strongly reshaped distribution. The agreement between calculation and experiment confirms that the measured deformation follows from the propagation-induced change in the relative phases of the spatiotemporal spectral components.

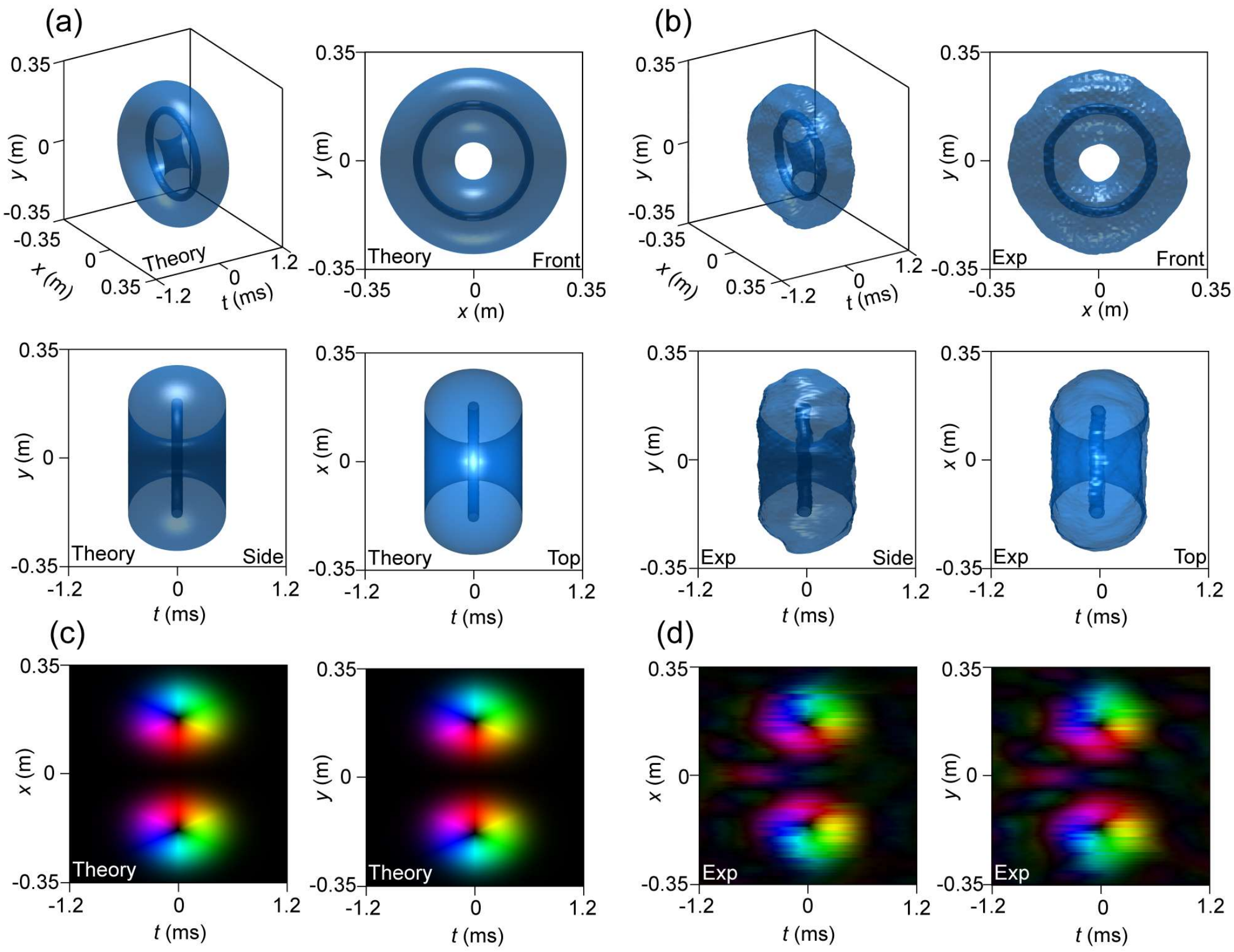

**FIG. 2. Experimental observation of a scalar acoustic toroidal vortex.**

(a) Theoretical iso-amplitude surface of the scalar acoustic toroidal vortex, viewed from different perspectives. The corresponding front, side, and top views reveal the toroidal structure and the closed phase singularity ring. (b) Experimentally reconstructed iso-amplitude surface of the acoustic toroidal vortex. (c) Theoretical amplitude and phase distributions in the radial–time planes. (d) Measured radial–time distributions, showing the characteristic spiral phase structure and the phase singularity in the spatiotemporal cross sections.

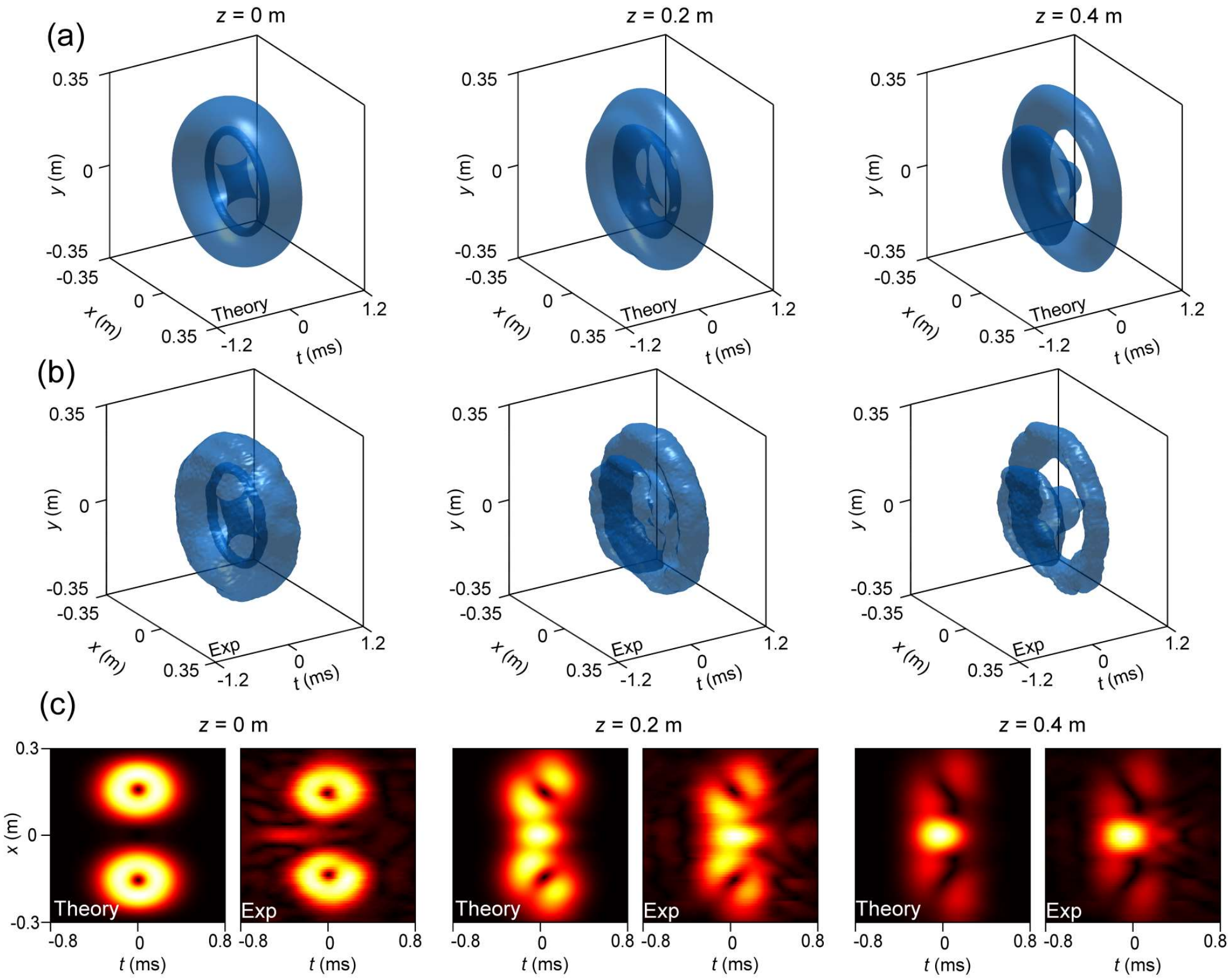

**FIG. 3. Propagation evolution of a scalar acoustic toroidal vortex.**

(a), (b) Theoretical and experimentally reconstructed iso-amplitude surfaces of the scalar acoustic toroidal vortex at propagation distances of $z = 0, 0.2, \text{and } 0.4\ \text{m}$, respectively, showing the evolution of the toroidal structure during propagation. (c) Theoretical and measured amplitude distributions in the radial–temporal planes at different propagation distances, revealing the corresponding spatiotemporal evolution of the vortex structure.

**Scalar acoustic hopfions from coupled toroidal and poloidal windings**

We now introduce a toroidal phase winding to organize the local vortex sections into a global phase texture. At a fixed toroidal position, the poloidal winding retains the local phase circulation around the vortex core. A nonzero toroidal winding adds a continuously varying phase offset as one moves around the closed ring. Consequently, a point of fixed phase shifts along the poloidal direction from one radial–temporal section to another, and tracing this point around the ring generates a closed phase fiber on the toroidal envelope. For the fundamental case $\left(l_{\text{tor}}, l_{\text{por}}\right) = (1,1)$, different phase fibers are mutually linked, forming a scalar acoustic hopfion.

Figure 4(a) presents the calculated scalar acoustic hopfion. The phase is encoded by color on a toroidal iso-amplitude surface, revealing the simultaneous phase variation along the toroidal and poloidal directions. Eight representative phase fibers, corresponding to eight equally spaced phase values over a $2\pi$ interval, are extracted from the field. These closed curves wind once around both the toroidal and poloidal directions and are mutually linked, forming the characteristic Hopf-linked configuration. The measured field yields eight closed phase fibers that wind around both toroidal cycles and remain mutually linked, directly revealing the Hopf-linked phase texture.

The coupled phase winding is further verified by the orthogonal sections shown in Figs. 4(c) and 4(d), where brightness represents the normalized pressure amplitude and hue encodes the phase. In the $x - t$ plane at $y = 0$ and the $y - t$ plane at $x = 0$, the singular ring is intersected at two pressure zeros. The phase circulates around each zero, revealing the poloidal winding inherited from the toroidal vortex. By contrast, the $x - y$ section at $t = 0$ exhibits a continuous azimuthal phase variation around the annular pressure distribution, directly confirming the additional toroidal winding. Taken together, the three-dimensional linked phase fibers and the complementary sectional phase circulations establish the simultaneous presence of toroidal and poloidal windings, and hence the formation of a scalar acoustic hopfion.

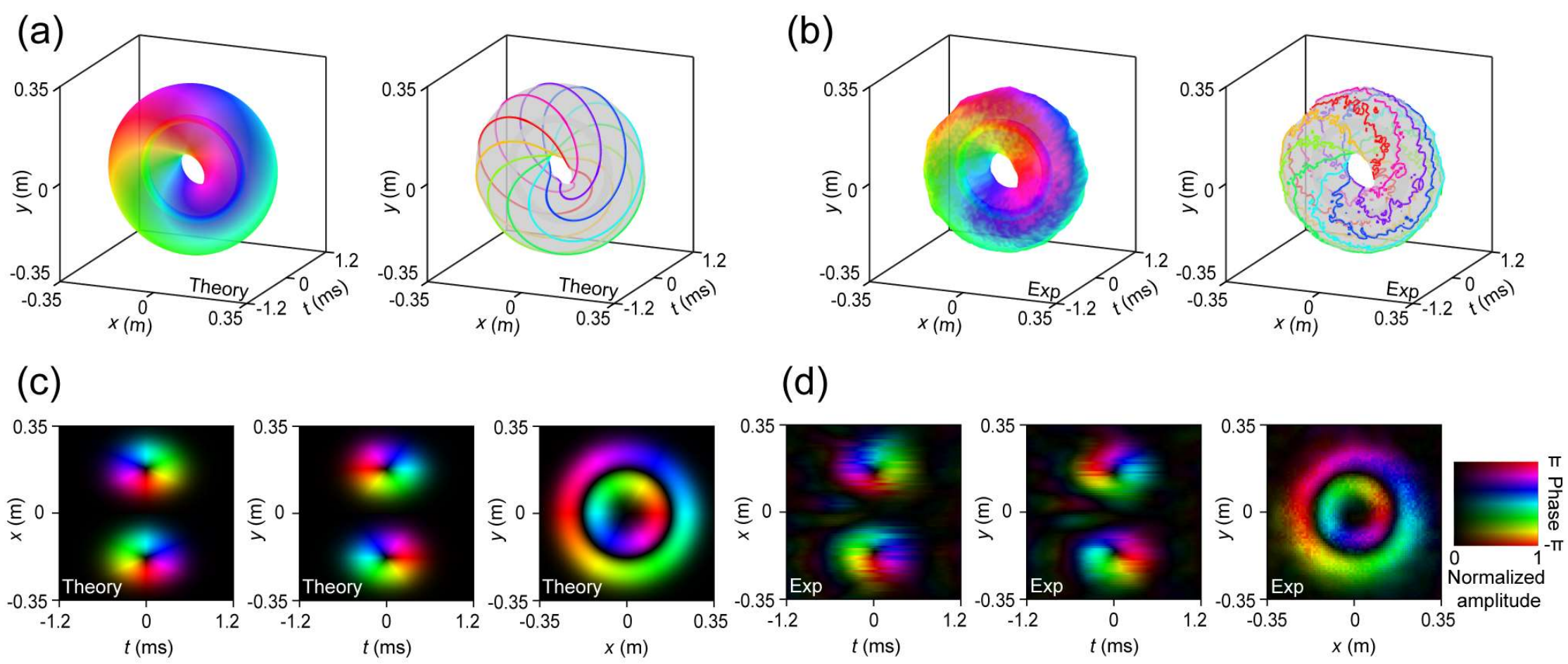

**FIG. 4. Observation of a scalar acoustic hopfion**

(a) Theoretical iso-amplitude surface and equiphase lines of the scalar acoustic Hopfion. The colored closed loops represent different phase fibers on the toroidal structure. (b) Experimentally reconstructed iso-amplitude surface and equiphase lines. (c), (d) Theoretical and measured amplitude-phase distributions in the transverse spatial and radial–temporal planes, respectively, revealing the coupled toroidal and poloidal phase windings.

**Programmable torus knots and links with independently tunable windings**

The topology of a scalar acoustic hopfion can be directly visualized by tracing its phase fibers, which are closed curves of constant acoustic phase on a toroidal iso-amplitude surface. For the field introduced above, the phase distribution on this surface can be expressed as

$$\Phi(\theta, \varphi) = l_{\mathrm{tor}}\varphi - l_{\mathrm{pol}}\theta, \tag{7}$$

where $\varphi$ and $\theta$ are the toroidal and poloidal angular coordinates, respectively. The integers $l_{\mathrm{tor}}$ and $l_{\mathrm{pol}}$ therefore provide independent control over the two phase windings. For a selected phase value $\Phi_j$, the condition

$$l_{\mathrm{tor}}\varphi - l_{\mathrm{pol}}\theta = \Phi_j \tag{8}$$

defines one or more closed curves on the toroidal envelope. Figure 5 shows two representative phase fibers corresponding to $\Phi = 0$ and $\Phi = \pi$, plotted in cyan and red, respectively, for six combinations of $(l_{\mathrm{tor}}, l_{\mathrm{pol}})$. The translucent gray surface indicates the underlying toroidal iso-amplitude envelope.

The Hopf invariant is given by the product of the two winding numbers,

$$Q_H = l_{\text{tor}} l_{\text{pol}}. \tag{9}$$

For two distinct phase values $\Phi_a$ and $\Phi_b$, $Q_{\text{H}}$ gives the total algebraic linking number between the corresponding complete phase fibers $C_{\Phi_a}$ and $C_{\Phi_b}$. When either fiber contains multiple disconnected components, this total is understood as the sum of the pairwise Gauss linking numbers over all component pairs. Accordingly, $|l_{\text{tor}} l_{\text{pol}}|$ determines the total number of algebraic linkings. This quantity is independent of the selected phase values and remains invariant under smooth deformations that do not involve cutting or reconnection.

The number of connected components in each phase fiber is determined by

$$d = \gcd(|l_{\text{tor}}|, |l_{\text{pol}}|). \tag{10}$$

where gcd denotes the greatest common divisor. Each phase fiber therefore decomposes as

$$C_{\Phi_j} = \bigcup_{\alpha=1}^{d} C_{\Phi_j}^{(\alpha)}. \tag{11}$$

We further define the reduced winding numbers as $l'_{\text{tor}} = |l_{\text{tor}}|/d$, $l'_{\text{pol}} = |l_{\text{pol}}|/d$, which are coprime by construction. Every component winds $l'_{\text{tor}}$ times in the poloidal direction and $l'_{\text{pol}}$ times in the toroidal direction. Coprime winding numbers $(d = 1)$ produce a single connected phase fiber. If both reduced winding numbers exceed unity, this fiber forms a nontrivial torus knot. For noncoprime winding numbers $(d > 1)$, each phase fiber splits into exactly $d$ disconnected components, with each component characterized by the corresponding reduced winding numbers. In the ideal toroidal geometry, the pairwise linking number between one component of $C_{\Phi_a}$ and one component of $C_{\Phi_b}$ is $l_{\text{tor}} l_{\text{pol}}/d^2$, and summation over the $d^2$ component pairs recovers $Q_{\text{H}} = l_{\text{tor}} l_{\text{pol}}$.

These relations are directly demonstrated in Fig. 5, where the cyan and red curves represent the phase fibers at $\Phi = 0$ and $\Phi = \pi$, respectively. For $(l_{\mathrm{tor}}, l_{\mathrm{pol}}) = (1,1)$, each phase value produces one closed loop, and the red and cyan loops form a fundamental Hopf link with $Q_{\mathrm{H}} = 1$ [Fig. 5(a)]. Increasing either winding number to two generates the (1,2) and (2,1) states shown in Figs. 5(b) and 5(c). Since the two winding numbers remain coprime, each phase fiber is still a single closed curve, while the two fibers have a total linking number $|Q_{\mathrm{H}}| = 2$. Although these states share the same linking number, their trajectories are geometrically distinct because the winding is distributed differently between the toroidal and poloidal cycles. A qualitatively different structure appears for (2,2), as shown in Fig. 5(d). In this case, $d = 2$, and each selected phase fiber separates into two disconnected loops. Each red component is linked once with each cyan component, yielding four red–cyan component pairs and a total linking number $|Q_{\mathrm{H}}| = 4$.

For the coprime pair (3,2), each phase value again produces a single connected curve. Because both reduced winding numbers exceed unity, the resulting phase fiber is a nontrivial torus knot, topologically equivalent to a trefoil knot. The two selected fibers have a total linking number $|Q_{\mathrm{H}}| = 6$, as illustrated in Fig. 5(e). By contrast, the (5,1) phase fiber remains topologically unknotted because one of its reduced winding numbers is unity, despite its pronounced fivefold winding geometry. The red and cyan fibers are linked five times, consistent with $|Q_{\mathrm{H}}| = 5$, as shown in Fig. 5(f).

The six configurations further separate several distinct properties of the phase-fiber topology. The (1,2) and (2,1) states have the same total linking number but follow different phase-fiber trajectories because the two windings are assigned to physically inequivalent cycles, one purely spatial and the other radial–temporal. The (2,2) state changes the connectivity of each complete phase fiber, whereas the (3,2) state produces nontrivial self-knotting in the form of a trefoil. By contrast, the strongly wound (5,1) fiber remains topologically unknotted. Thus, geometrical winding, mutual linking, connectivity, and self-knotting represent distinct characteristics of the scalar phase texture and are not fully captured by the total linking number alone. The reconstructed fibers experimentally resolve these distinctions across all six configurations, demonstrating programmable control through the two independently tunable winding numbers.

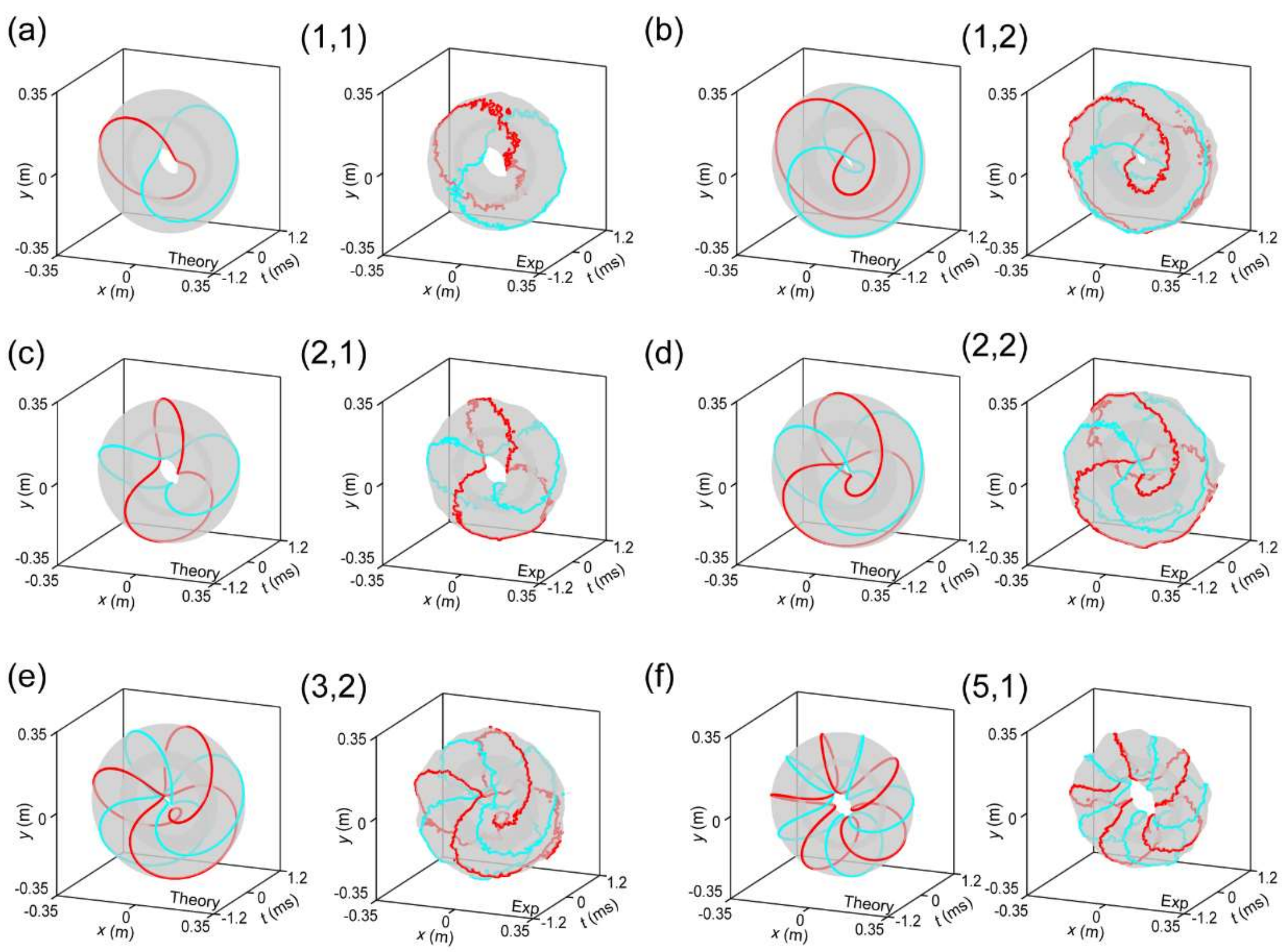

**FIG. 5. Programmable scalar acoustic hopfions with tunable winding numbers.** (a)–(f) Theoretical and experimentally reconstructed equiphase lines for scalar acoustic hopfions with different toroidal and poloidal winding numbers: $(l_{\mathrm{tor}}, l_{\mathrm{pol}}) =$ $(1,1), (1,2), (2,1), (2,2), (3,2),$ and $(5,1)$. The cyan and red curves represent the phase fibers at $\Phi = 0$ and $\Phi = \pi$ , respectively, whose linking topology is governed by the winding numbers.

## Discussion

In summary, we have realized freely propagating scalar acoustic toroidal vortices and reconstructed their full complex pressure fields in space and time. Their free-space evolution shows that geometric closure does not imply propagation invariance. Over the present bandwidth, transverse diffraction in air is not accompanied by comparable material group-velocity dispersion, whereas anomalous group-velocity dispersion in engineered optical media can allow the radial and temporal profiles to evolve on comparable scales [20]. Introducing an independently controlled toroidal winding extends the elementary vortex to scalar acoustic hopfions with programmable phase-fiber topology. The poloidal winding defines the local phase circulation around the vortex core, while the toroidal winding determines how equal-phase points in different radial–temporal sections connect around the closed ring. These two windings organize the local vortex structure into global phase fibers with controllable linking, connectivity, and knotting. The resulting configurations further show that total linking alone does not fully characterize the phase texture, since identical linking can arise from different winding allocations, while pronounced geometrical winding does not necessarily imply nontrivial knotting. Programmable synthesis and full complex-field reconstruction provide direct access to both the propagation dynamics of the toroidal wave packet and the topology of its phase fibers. This capability may enable further studies of deformation, interaction, and reconnection in toroidal wave fields [50–53], while offering additional winding degrees of freedom for acoustic encoding and structured wave–matter interactions.

## Acknowledgements

This work was supported by the National Key R&D Program of China (Grant No. 2023YFA1406904), the National Natural Science Foundation of China (Grants No. 52250363 and 52203358), the Natural Science Foundation of Jiangsu Province (Grants No. BK20232048 and BK20233001).

**Data availability**

The data that support the findings of this article are not publicly available. The data are available from the authors upon reasonable request.